\input psfig.sty

\documentclass{article}
\begin{document}

\title{Pulsar Astronomy: the HST Contribution}

\author{P.A. Caraveo$^1$ \and R. P. Mignani$^2$ \and G.G.Pavlov$^3$ \and G.F. Bignami$^4$}
\date{ }

\maketitle

\noindent
$^1$ IFC-CNR,  Via Bassini  15,   I-20133 Milan, Italy

\noindent
$^2$ ST-ECF, Karl Schwarzschild  Str.2, D8574O   Garching   b.  Munchen, Germany

\noindent
$^3$ Pennsylvania State Univ., 525 Davey Lab, University Park, PA 16802, USA

\noindent
$^4$ ASI, Via Liegi 26, I-00198 Rome, Italy

\begin{abstract}
HST observations have contributed significantly to our knowledge on the behaviour of Isolated Neutron 
Stars (INSs) as optical emitters.  First, HST has been instrumental both to discover new optical counterparts 
(PSR B1055-52, PSR B1929+10, PSR B0950+08) and to confirm proposed identifications (PSR 
B0656+14). Second, HST multicolor photometry provided useful information to characterize the optical 
emission mechanism(s) at work in middle-aged INSs like PSR B0656+14 and Geminga. Last, but not least, 
the superior angular resolution of the HST allowed both to perform a very accurate morphological study of  
the plerionic environments of young pulsars (e.g. the Crab and PSR B0540-69) and to perform very accurate 
astrometric measurements yielding proper motions  (Crab, Vela, Geminga, PSR B0656+14) and parallaxes 
(Geminga). 
\end{abstract}

\section{Introduction}
Although conspicous INSs such as the Crab and Vela pulsars have been observed from the very beginning of 
the mission,  HST started to play a key role on the study of  the optical behaviour of these faint targets after 
the first refurbishing  mission in 1993.  The study did not proceed systematically, e.g.  from the brighter to 
the dimmer, but rather following a random walk dictated by the allocation of observing time. Table 1 lists all 
the INSs (be they bona fide pulsars or radio-silent neutron stars) observed so far by the HST.

\begin{table}
\begin{tabular}{ccccc} \hline \hline \\
ID & Log(yr) & Log(dE/dt) & D(kpc) & mag \\ \hline
\multicolumn{5}{c}{\it{Bona fide Pulsars}} \\ \hline
Crab & 3.1 & 38.6 & 2.0 & 16.6 \\
B0540-69 & 3.2 & 38.2 & 55 & 22.5 \\
Vela & 4.1 & 36.8 & 0.5 & 23.6 \\
B0656+14 & 5.0 & 34.6 & 0.76 & 25.0 \\
Geminga & 5.5 & 34.5 & 0.16(\dag) & 25.5 \\
B1055-52 & 5.7 & 34.5 & 1.5 & 24.9(U) \\
B1929+10 & 6.5 & 33.6 & 0.17(\ddag) & 25.7(U) \\
B0950+08 & 7.2 & 32.7 & 0.28(\ddag) & 27.1(U) \\ \hline
\multicolumn{5}{c}{\it{Radio-silent INSs}} \\ \hline
RXJ 1856-3754 & & &  $<0.13$ & 25.6 \\ \hline \hline
\end{tabular}
\caption{(\dag) determined from HST parallax; (\ddag) determined from radio parallax.
Isolated Neutron Stars observed so far by the HST. Two more objects (PSR B1509-58 and RXJ0720-
3125) with a proposed optical ID have been studied from the ground only. The table lists the neutron stars' 
ID (first column), their spin-down age (column two), their rotational energy loss in erg/s (column three), 
their nominal radio distance in kpc (column four) and their magnitude in the V band, unless otherwise 
indicated (column five). Horizontal lines separate decades of pulsar spin-down age.}   
\end{table}

Although their number is limited, the objects in Table 1 sample 10 magnitude in brightness and 4 decades in 
age, going from the youngest pulsars, such as the  Crab and PSR B0540-69, to rather old ones, such as PSR 
B0950+08.

All INSs, but the Crab, are faint. All challenging, sometimes plainly impossible to observe from the ground. 
This was the case of PSR B1055-52 (Mignani et al. 1997\cite{7}) which,  together with PSR B1929+10 and PSR 
B0950+08 (Pavlov et al. 1996\cite{10}) have been seen only with the  HST using the FOC and the U filter. To the 
score of HST identifications we can add the INS candidate RXJ 1856-3754 (Walter \& Matthews 1997\cite{14}).

\section{The Data}
Over the years, HST has collected light curves, for the  Crab (Percival et al. 1993\cite{13}) and PSR B0540-69 (Boyd 
et al. 1995\cite{2}), spectra, for the same two objects (Gull et al. 1998\cite{4}; Hill et al. 1997\cite{5}), and images in different 
filters for all of them. The major results obtained by HST in pulsar astronomy have been reviewed by 
Mignani et al. (2000)\cite{9}. The observational efforts pursued by different groups with the imaging  instruments 
on board HST are summarized in Table 2, where, for sake of clarity, the spectral coverage provided by HST 
has been roughly divided in two infrared channels (IR and I), four optical ones (R,V,B,U)- plus narrow 
bands (NB)- and one ultraviolet. In Table  2, NICMOS, WFPC2, and  FOC observations are indicated. If an 
observation has been done more than once,  the number in parenthesis gives the number of repetitions.

\begin{table}
\small
\begin{tabular}{lcccccccc} \hline \hline \\
ID & IR & I & R & V & B & U & UV & NB \\ \hline
Crab & & & & & & & & 547M (several) \\
B0540-69 & & & & WFPC2 & & & & 656N, 658N \\ \hline
Vela & & WFPC2 & WFPC2 & WFPC2(5) & & & & \\ \hline
B0656+14 & NICMOS & & & WFPC2(2) & FOC & FOC & FOC & \\
Geminga & NICMOS & & WFPC2 & WFPC2(4) & FOC & FOC & FOC & \\
{\it B1055-52} & & & & & & {\it FOC} & & \\ \hline
{\it B1929+10} & & & & & & {\it FOC} & {\it FOC} & \\
{\it B0950+08} & & & & & & & {\it FOC} & \\ \hline
{\it RXJ 1856-3754} & & & & {\it WFPC2(2)} & {\it WFPC2} & {\it WFPC2(2)} & {\it WFPC2} \\ \hline \hline
\end{tabular}

\caption{Summary of the multicolor photometry of INSs obtained by the HST. Objects are separated as in 
Table 1. {\it Italic} indicates the pulsars first observed with the HST and the detections wavebands.}

\end{table}
\normalsize

Table 2 shows quite eloquently that not all the entries in Table 1 received the same amount of observing 
time: it is worth noticing that, apart from  the "dancing Crab", the objects with the highest number of 
observations is the rather dim Geminga, followed by PSRB0656+14, to show that objects fainter than V=25 
were not discriminated in this study. The amount of information contained in this comprehensive list has 
been used:
\begin{itemize} 
\item{to measure pulsars' proper motions and parallactic displacements,} 
\item{to study plerion phenomenology}
\item{to assess the spectral distribution of objects too faint for spectroscopy}
\end{itemize}
The major achievements are summarized in the next sections.

\begin{figure}
\centerline{\hbox{\psfig{figure=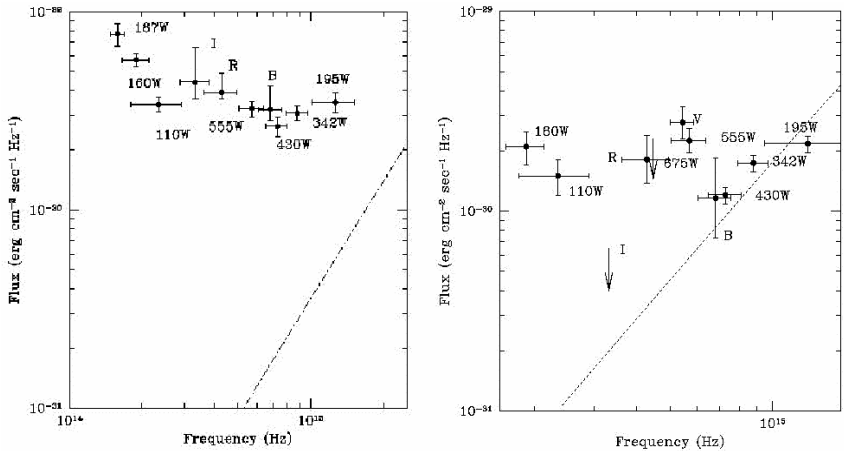,height=6cm,clip=}}}  
\caption{Optical flux distribution of PSRB0656+14 (left) and Geminga (right) as obtained (see Koptsevich et al. 2000) from the  
combination of ground-based and HST photometry (three digits labels). While the two objects belong to the  
same class of middle-aged pulsars with similar energetics, their optical properties appear different. Dotted  
lines represents the extrapolation of the soft X-ray black-body emission
measured by ROSAT. }  
\end{figure}

\section{High Resolution Imaging}

\subsection{INS astrometry.}
For all the pulsars observed more than once, namely the Crab, Vela, PSR B0656+14 and Geminga, a proper 
motion has been measured, yielding also new and independent measurements of the objects' transverse 
velocities. This topic is reviewed in these proceedings by Mignani et al. Sometimes, the accurate 
determination of the proper motion has been a by-product of a sequence of observations aimed at the 
measurement of the object's parallactic displacement and hence its distance (see also De Luca et al., these 
proceedings). This has been done for Geminga (Caraveo et al. 1996\cite{3}) and is currently underway for the Vela 
pulsar. Determining the distance to a pulsar allows the assessment of the absolute optical luminosity which, 
compared with the overall energy loss dE/dt, yields the efficiency to convert rotational energy loss into 
optical emission, an important parameter in pulsar modelling.

\subsection{Morphology studies}
HST imaging of Crab, Vela and PSR B0540-69 allows one to trace the relativistic pulsar wind and to better 
study the plerion phenomenology. Moreover, with the proper motion vectors clearly aligned  with the axes 
of symmetry of the Crab and Vela plerions, proper motions, or rather the mechanisms responsible for  them,  
seem to play a role in shaping the inner remnants (see Mignani et al., these  proceedings, and Pavlov et al. 
2000\cite{12}).

Comparisons between HST frames and recently obtained Chandra high resolution images open new avenues 
to study the multiwavelength behaviour of young  energetic plerions. The case of PSR B0540-69 is 
discussed in an accompanying paper by Caraveo et al. .

\subsection{Multicolor Imaging}
HST multicolor imaging appears to be the next best thing to a spectrum for studying the spectral shape of 
faint objects and discriminating between thermal emission from the INS surface and non thermal 
magnetospheric one.

Two classical examples are 
\begin{itemize}
\item{PSR B0656+14, where Pavlov et al. (1997)\cite{11} have shown a composite spectral shape featuring both a 
thermal and non-thermal components (see Fig.1, left panel)}
\item{Geminga, for which Bignami et al. (1996)\cite{1} and Mignani et al. (1998)\cite{8} have provided the evidence of a 
cyclotron spectral feature on top of  a thermal continuum (see Fig.1, right panel). If correct, the 
cyclotron identification (discussed also by Jacchia et al. 1999\cite{6}) of the feature would provide the first in 
situ measurement of the magnetic field of an isolated neutron star.} 
\end{itemize}

\section{Conclusions.}
All in all, the study of INSs, in spite of their faintness, has yielded a wealth  of interesting results definitely 
worth the time and efforts devoted to them. New identifications have been secured while new insights have 
been achieved for pulsars already identified.

Of course, a lot remains to be done. Young pulsars are definitely promising targets, thus we should 
concentrate on newly discovered young objects, such as the 16 msec one in the LMC. Here the timing 
capability of the STIS, so far poorly exploited, should be fully used.

Radio quiet candidate neutron stars are also promising, although admittedly demanding, targets.

\end{document}